\pdfoutput=1
\documentclass{article}

\usepackage[final]{neurips_2022}

\usepackage[utf8]{inputenc} 
\usepackage[T1]{fontenc}    
\usepackage{hyperref}       
\usepackage{url}            
\usepackage{booktabs}       
\usepackage{amsfonts}       
\usepackage{nicefrac}       
\usepackage{microtype}      
\usepackage{xcolor}         
\usepackage{amsmath,amsthm,amssymb}
\usepackage{mathrsfs}
\usepackage{todonotes}
\usepackage{graphicx}
\usepackage{pdflscape}
\usepackage{algorithm}
\usepackage{algpseudocode}

\title{An efficient graph generative model for navigating ultra-large combinatorial synthesis libraries}

\author{%
  Aryan Pedawi \\
  Atomwise Inc. \\
  \texttt{aryan@atomwise.com} \\
  \And
  Paweł Gniewek \\
  Atomwise Inc. \\
  \texttt{pawel@atomwise.com} \\
  \And
  Chaoyi Chang \\
  Atomwise Inc. \\
  \texttt{cchang373@atomwise.com} \\
  \AND
  Brandon M. Anderson\thanks{Work performed while at Atomwise Inc. Current affiliation is with Atomic AI.}\;\,\thanks{Equal senior contributions.} \\
  Atomwise Inc. \\
  \texttt{branderson@gmail.com} \\
  \And
  Henry van den Bedem${}^\dagger$ \\
  Atomwise Inc. \\ UCSF, Dept. of Bioengineering \& Therapeutic Sciences \\
  \texttt{vdbedem@atomwise.com}
}

\begin{document}

\maketitle

\begin{abstract}
Virtual, make-on-demand chemical libraries have transformed early-stage drug discovery by unlocking vast, synthetically accessible regions of chemical space. Recent years have witnessed rapid growth in these libraries from millions to trillions of compounds, hiding undiscovered, potent hits for a variety of therapeutic targets. However, they are quickly approaching a size beyond that which permits explicit enumeration, presenting new challenges for virtual screening. To overcome these challenges, we propose the \textbf{C}ombinatorial \textbf{S}ynthesis \textbf{L}ibrary \textbf{V}ariational \textbf{A}uto-\textbf{E}ncoder (\textbf{CSLVAE}). The proposed generative model represents such libraries as a differentiable, hierarchically-organized database. Given a compound from the library, the molecular encoder constructs a query for retrieval, which is utilized by the molecular decoder to reconstruct the compound by first decoding its chemical reaction and subsequently decoding its reactants. Our design minimizes autoregression in the decoder, facilitating the generation of large, valid molecular graphs. Our method performs fast and parallel batch inference for ultra-large synthesis libraries, enabling a number of important applications in early-stage drug discovery. Compounds proposed by our method are guaranteed to be in the library, and thus synthetically and cost-effectively accessible. Importantly, CSLVAE can encode out-of-library compounds and search for in-library analogues. In experiments, we demonstrate the capabilities of the proposed method in the navigation of massive combinatorial synthesis libraries.
\end{abstract}

\section{Introduction}

Virtual high throughput screening (vHTS) \cite{Shoichet2004} has gained significant traction in early-stage drug discovery, owing in no small part to make-on-demand chemical libraries utilizing a \emph{combinatorial synthesis} construction. These combinatorial synthesis libraries (CSLs) enable access to ultra-large swaths of chemical space from a considerably smaller set of chemically accessible building blocks that can be combined according to known synthesis routines. In recent years, these libraries have grown from millions, to billions, and now to trillions of compounds \cite{Irwin2016, Lyu2019, sadybekov2022synthon, Warr2022}. For example, the Enamine REadily AccessibLe (REAL) libraries \cite{grygorenko2020generating} leverage off-the-shelf molecular building blocks and parallel synthesis, permitting lead times on the order of a few weeks and ushering in an era of ever-decreasing latency between \emph{in silico} and \emph{in vitro} high throughput screening.

As a result of the combinatorial explosion these constructions enable, early-stage drug discovery has now ``crossed the Rubicon'' into the non-enumerative regime. This presents new challenges for \emph{in silico} hit discovery and optimization, which  rely  on screening explicitly enumerated compounds. These methods are ill suited to the non-enumerative setting, scaling linearly with the number of compounds, which motivates our interest in designing scalable approaches for navigating such libraries.

In this paper, we propose the \textbf{C}ombinatorial \textbf{S}ynthesis \textbf{L}ibrary \textbf{V}ariational \textbf{A}uto-\textbf{E}ncoder, or \textbf{CSLVAE} for short (pronounced like \emph{c'est la vie}). CSLVAE is a graph-based generative model that exploits the structure of CSLs towards efficient navigation of the relevant chemical space. Our model learns a hierarchy of keys over the components of the library, and uses these keys to process queries for retrieval. The encoder processes a molecular graph and returns as output a query vector, which the decoder uses to retrieve the molecule through an efficient sequence of query-key comparisons that utilizes the hierarchical construction of CSLs, requiring minimal autoregression and admitting efficient parallelization.

The main contributions of this paper are as follows:
\begin{itemize}
\item We present a novel graph generative model which acts as a ``neural database,'' providing random access to ultra-large, non-enumerable compound libraries. As such, our model is guaranteed to generate valid and cost-effectively accessible molecules.
\item Our model overcomes challenges with long autoregressive chains in compound generation, improving scalability to large molecular graphs.
\item Our model reduces the number of parameters ten-fold relative to comparable methods, and offers considerable improvements in computational complexity for searching through CSLs.
\end{itemize}

\section{Related Work}

\paragraph{Virtual high throughput screening and enumeration}

Often, the first step in a vHTS campaign is preparing a compound library for subsequent use \cite{acharya2020supercomputer,gorgulla2020open}. While highly optimized compound sampling and scoring techniques have been developed \cite{graff2021accelerating}, these approaches nevertheless rely on an exhaustively accessible library. An exception is the virtual synthon hierarchical enumeration screening (V-SYNTHES) approach \cite{sadybekov2022synthon}, which leverages the modular nature of parallel synthesis libraries. However, by design, V-SYNTHES does not permit query-based random access. On the other hand, SpaceMACS \cite{schmidt2021maximum} and SpaceLight \cite{bellmann2020topological} can provide query-based access to modular libraries by decomposing the query into fragments, and matching those by similarity search to synthons in the library. In parallel to these efforts, machine learning has received significant attention in vHTS: for predicting activity scores given docked conformations \cite{gniewek2021learning,ragoza2017protein,wallach2015atomnet}, predicting activity scores given a ligand and protein separately (undocked) \cite{ozturk2018deepdta,tsubaki2019compound}, and in improving or altogether replacing classical molecular docking with machine learning approaches \cite{paggi2021leveraging,stafford2022atomnet,stark2022equibind}.

\paragraph{Deep learning approaches to molecular generation}

De novo drug design has assumed an increasingly prominent role in identifying novel chemical matter in drug discovery campaigns \cite{Chen2021a,Meyers2021,Sousa2021,Tong2021}. The two dominant neural network-based paradigms for molecular generation are text-based and graph-based generative models. Early work in text-based generative models (also called chemical language models) applied recurrent neural networks to SMILES strings \cite{gomez2018automatic,segler2018generating}. Although these methods have shown a great deal of promise and spurred interest in molecular generation within the ML community, they are not guaranteed to produce valid SMILES strings. Approaches utilizing grammar constraints of SMILES notation have been proposed to improve validity \cite{dai2018syntax,kusner2017grammar}; separately the recently proposed SELFIES notation \cite{krenn2020self,nigam2021beyond} guarantees validity and has seen increased adoption as such. In both cases, however, there remain known drawbacks to modeling with such text-based representations of chemical matter (e.g., surjectivity, similar molecular structures having possibly large edit distances).

For some applications, it is of interest to utilize generative models which can fit molecular databases, permitting the navigation of these databases via the fitted model. The ability of language models to fit molecular databases was investigated in a prior study \cite{arus2019exploring} that applied deep language models to GDB-13 \cite{blum2009970}, a database of $10^9$ compounds formed by fully enumerating molecules up to 13 atoms of element types C, N, O, S, and Cl, subject to simple chemical stability and synthetic feasibility rules. The authors trained on 0.1\% of the total library and find that the model was capable of covering roughly 70\% of compounds in the GDB-13 library. Furthermore, the language model they trained generates compounds not satisfying the GDB-13 construction in approximately 15\% of cases.

Graph generative models have received significant attention in recent years as an alternative to their text-based counterparts. The earliest of these models focused on generating graphs of a constant size in a single shot \cite{simonovsky2018graphvae} or generating graphs of arbitrary size autoregressively, one atom or bond at a time \cite{liu2018constrained,Popova2019,You2018,Mercado2021}. These approaches also struggle to reliably produce chemically valid molecules and encounter difficulties with large molecular graphs.

In an effort to address both points, fragment-based graph generative models have been proposed and are growing in popularity \cite{jin2018junction,Jin2020,jin2020multi,Kong2021}. These models have the advantage of guaranteeing chemical validity by decomposing molecules into valid sub-components and explicitly disallowing actions which yield invalid combinations of fragments. Such explicit validity checks can be performed on every action, at the cost of additional computation. While other text- and graph-based generative models tend to struggle with large molecular graphs due to the long autoregression chains needed to produce them, fragment-based graph generative models require autoregression lengths on the order of the number of fragments that comprise a molecule. This can be significant when the fragments themselves contain many atoms. 

However, some issues persist due to general difficulties with autoregressive graph generation. Unlike text-based models, in which there is less ambiguity in the autoregression order (e.g., tokens are typically decoded in left-to-right order), graphs have no such canonical node order, which presents challenges for graph-based autoencoders \cite{liu2018constrained,winter2021permutation}. Furthermore, although they require shorter autoregression chains than their counterparts, existing fragment-based graph generative models nevertheless require autoregression lengths that grow in the overall size of the molecule since autoregressive decoding cannot be effectively parallelized.

While fragment-based graph generative models and SELFIES-based language models each address the issue of chemical validity, there is the separate challenge of synthetic accessibility. Prior work has cast doubt on the synthetic feasibility of compounds proposed by many existing generative models \cite{Gao2020}, which can limit the practical utility of these models in drug discovery applications if not appropriately addressed. Subsequent work has attempted to improve on these shortcomings by (i) including explicit penalties for synthetic inaccessibility via a scoring function \cite{Griffiths2020}, (ii) limiting the model to fragments from known compounds \cite{maziarz2021,Polishchuk2020,Takeuchi2021}, or (iii) inducing bias towards simple and known synthetic pathways \cite{Bradshaw2019,Bradshaw2020,Horwood2020}.

\begin{figure}
    \centering
    \includegraphics[origin=c,width=5.5in]{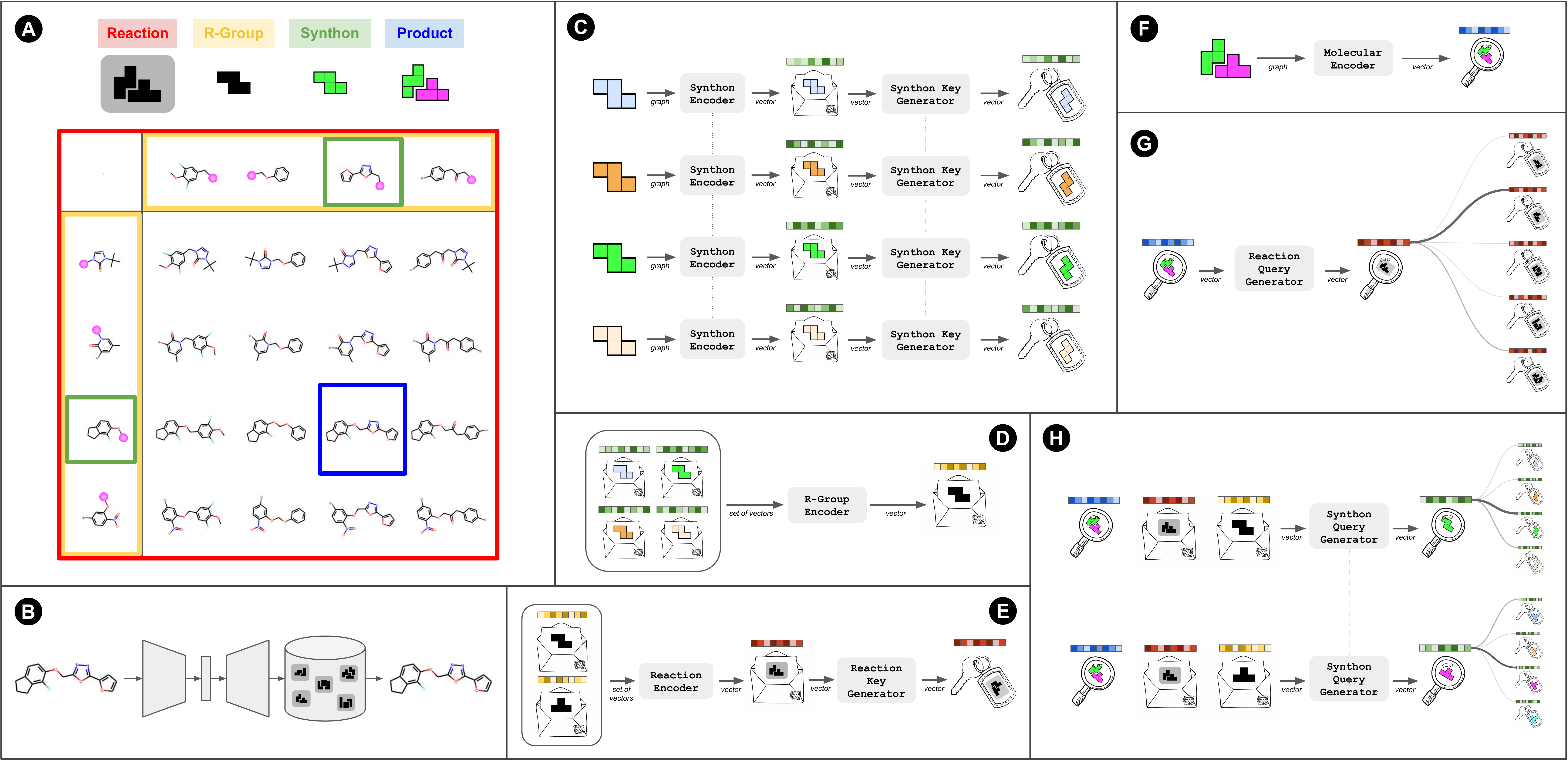}
    \caption{\textbf{Overview of CSLVAE.} \textbf{Panel A:} Illustration of a combinatorial synthesis table, which form the basis of combinatorial synthesis libraries (CSLs). \textbf{Panel B:} CSLVAE is an autoencoder whose encoder takes molecular graphs as input and returns query vectors as output, which the decoder then utilizes to retrieve the corresponding products from the library. \textbf{Panels C-E:} The CSLVAE library encoder represents CSLs via a hierarchy of learned representations. It proceeds by encoding individual synthons, represented as molecular graphs, with a graph neural network. It subsequently encodes R-groups with a set neural network over their constituent synthon representations. Finally, it encodes reactions with a set neural network over their constituent R-group representations. \textbf{Panel F:} The molecular encoder returns a query from an input molecular graph (in this case, a product from the library). \textbf{Panels G-H:} The molecular decoder processes the query to retrieve a product from the CSL, by first decoding the reaction type and subsequently decoding one synthon per R-group via query-key lookups.}
    \label{figure1}
\end{figure}

\section{Methodology}

This section formalizes combinatorial synthesis libraries and the proposed model, CSLVAE. Figure \ref{figure1} provides an illustration of the approach. Details on the architectures used for the various modules which comprise CSLVAE can be found in the \hyperref[sec:appendix]{Appendix}.

\subsection{Preliminaries}

CSLs are composed of a set of multi-dimensional synthesis tables (Figure \ref{figure1}, Panel A). Each synthesis table describes a multi-component chemical \emph{reaction}, which we denote by an index $t\in T$ (e.g., a natural number). Following \cite{sadybekov2022synthon}, we use the term \emph{R-group} to refer to a singular component in a reaction, and denote it by an index $r\in R$. A multi-component reaction takes several reactants (components) as input to produce a single molecule via chemical synthesis. We let $\psi: T\to\mathcal{P}(R)$ be the function which returns the set of R-groups $\psi(t)\subset R$ associated with a reaction $t$, where $\mathcal{P}(\cdot)$ denotes the power set function.

Each R-group is spanned by a possibly large number of molecular building blocks, called \emph{synthons}, which can be utilized in the corresponding reaction. We denote a synthon by an index $s\in S$ and represent it with a molecular graph, $\mathcal{G}_s\in\mathscr{G}_S$. We note that a synthon can belong to multiple R-groups. For convenience, we use $\sigma: R\to \mathcal{P}(S)$ to represent the function which returns the set of synthons $\sigma(r)\subset S$ belonging to a particular R-group $r$.

A \emph{product} $x\in X$ is a molecule that is synthesized according to a $k_t$-component reaction with R-groups $\psi(t)=(r_t^{(1)},\ldots,r^{(k_t)}_t)$ and a corresponding synthon tuple $u=(s^{(1)}_u,\ldots, s^{(k_t)}_u)$, where $s^{(i)}_u\in \sigma(r^{(i)}_t)$ for all $i=1,\ldots,k_t$. We define $f$ to be the synthesis rule which generates a compound $x$ from a reaction and synthon tuple pair $(t, u)$, i.e., $x := f(t, u)$. In short, as a simple analogy, one can think of a synthesis rule $t$ as specifying an equation of $k_t$ terms, where each term is an R-group, and their associated synthons correspond to the allowed values for the corresponding term.

Hence, a combinatorial synthesis library $\mathcal{D}\equiv(T,R,S,f,\psi,\sigma)$ is fully characterized by its reactions $T$, R-groups $R$, and synthons $S$, together with the synthesis rule $f$, reaction to R-groups mapping $\psi$, and R-group to synthons mapping $\sigma$. 

Using the language of probability, one can describe a distribution over products induced by $\mathcal{D}$ via the following factorization:
\begin{align}
p(x|\mathcal{D}) = \sum_{t\in T}\sum_{u\in U_t} p(t|\mathcal{D})\, p(u|t,\mathcal{D})\, p_f(x|t,u),\label{pdf}
\end{align}

where $U_t=\sigma(r_t^{(1)})\times\ldots\times \sigma(r_t^{(k_t)})$ is the set of all eligible synthon tuples for reaction $t$. This factorization describes the generative process in which one first samples a reaction $t\sim p(t|\mathcal{D})\propto |U_t|$, then samples a valid synthon tuple $u\sim p(u|t,\mathcal{D})=|U_t|^{-1}$ comprised of synthons from the respective R-groups in $t$, and joins these together via synthesis to form a product $x$ (via the deterministic rule $f$).

As written, all valid $(t,u)$ pairs in $\mathcal{D}$ are equally probable under $p$. Note that if every product in $\mathcal{D}$ can be reached according to just a single synthesis route, then $p(x|\mathcal{D})$ is a uniform distribution over the part of chemical space $X$ accessible by $\mathcal{D}$.

\subsection{Combinatorial Synthesis Library Variational Auto-Encoder (CSLVAE)}

We consider the task of looking up a product $x$ from a library $\mathcal{D}$, which amounts to finding the reaction $t$ and synthon tuple $u$ satisfying $x=f(t,u)$. This can be cast as an inference problem seeking $p(t, u|x,\mathcal{D})$. A latent variable model for $x$ gives rise to a variational formulation:
\begin{align}
p(t,u|x,\mathcal{D}) &= \int_{\bf z} p(t, u|{\bf z},\mathcal{D}) \, q({\bf z}|x)\, \text{d}{\bf z},
\end{align}

with $p({\bf z})$ denoting the prior distribution of the latent variable. One can further simplify the joint conditional distribution of $t$ and $u$ by first selecting the reaction $t$ and then selecting the synthons $s_u^{(1)},\ldots, s_u^{(k_t)}$ independently conditional on $t$:
\begin{align}
p(t,u|{\bf z},\mathcal{D}) &= p(t|{\bf z},\mathcal{D})\, p(u|{\bf z}, t,\mathcal{D}) \\
&= p(t|{\bf z}, \mathcal{D}) \prod_{i=1}^{k_t} p(s_u^{(i)} | {\bf z}, t,r^{(i)}_t,\mathcal{D}).\label{likelihood}
\end{align}

This gives rise to a strategy in which one encodes a molecule $x$ into the latent space, ${\bf z}\sim q({\bf z}|x)$, and proceeds by first decoding the reaction type $t\sim p(t|{\bf z},\mathcal{D})$ and then, conditional on the sampled reaction, decoding one synthon per R-group $s^{(i)}_u\sim p(s_u^{(i)}|{\bf z},t,r^{(i)}_t, \mathcal{D})$ for $i=1,\ldots, k_t$ to form the synthon tuple $u\in U_t$.

The resulting latent variable model is the proposed CSLVAE. Panels B-H in Figure \ref{figure1} give a step-by-step depiction of the method. In the forthcoming subsections, we will describe and formalize the three primary modules that comprise CSLVAE in detail: (i) the library encoder, (ii) the molecular encoder, and (iii) the molecular decoder. Module (ii) forms the basis for $q({\bf z}|x)$, while modules (i) and (iii) form the basis for $p(t,u|{\bf z},\mathcal{D})$.

\subsubsection{Library encoder}

A CSL $\mathcal{D}\equiv(T,R,S,f,\psi,\sigma)$ is organized hierarchically, with synthons $S$ at the bottom of the hierarchy, R-groups $R$ in the middle, and reactions $T$ at the top. We propose a strategy for learning an associated hierarchy of representations that describe the library at these three levels of resolution in an end-to-end fashion. These representations can then be used in retrieving results from queries into the library. The library encoder is illustrated in panels C-E of Figure \ref{figure1}.

We start from the bottom of the hierarchy with the synthons $S$. To learn a representation for each synthon in a manner that is fully inductive, we use a graph neural network to parameterize the $\texttt{SynthonEncoder}:\mathscr{G}_S\to\mathbb{R}^{d_S}$, which applies a sequence of message passing steps followed by a readout to arrive at a $d_S$-dimensional representation for each synthon.

Moving up the hierarchy, we use a deep set neural network to represent the R-groups $R$ by summarizing the representations for the synthons belonging to a particular R-group. Formally, the $\texttt{RGroupEncoder}: \mathbb{R}^{d_S}\times\cdots\times\mathbb{R}^{d_S}\to\mathbb{R}^{d_R}$ learns a $d_R$-dimensional representation for a given R-group from the set of $d_S$-dimensional representations of the constituent synthons. Following \cite{zaheer2017deep}, the R-group encoder has the form $\texttt{RGroupEncoder}(\{{\bf h}_s^S : \forall s\in \sigma(r)\}) = \rho\left(\bigoplus_{s\in \sigma(r)}\phi({\bf h}_s^S)\right)$, where $\phi$ and $\rho$ are parameterized by multi-layer perceptrons (MLPs) and $\bigoplus$ is a permutation-invariant aggregation operator. While other potentially more performant set-to-vector neural networks could be considered here, e.g., set transformers \cite{lee2019set}, an advantage of this simple construction is that it permits fast querying over library subsets at test time since $\phi({\bf h}_s^S)$ can be cached for all $s\in S$, thereby reducing the required computations for making queries into partitions of the library. We utilize mean pooling as the aggregation operator to focus on characteristics of the distribution of synthons in an R-group (as opposed to sum pooling which generally expresses characteristics of the multiset), which improves performance when dealing with R-groups of varying cardinality.

Finally, to represent reactions, we use yet another deep set neural network as the $\texttt{ReactionEncoder}: \mathbb{R}^{d_R}\times\cdots\times\mathbb{R}^{d_R}\to\mathbb{R}^{d_T}$. This module takes as input a variable-sized set of R-group representations corresponding to the reactants in $t$ and produces a $d_T$-dimensional representation for the reaction. We utilize sum as the aggregation operator to learn multiset properties of the R-groups in a reaction.

Putting these together, the representations cascade as follows:
\begin{align}
{\bf h}^S_s &= \texttt{SynthonEncoder}(\mathcal{G}_s), \label{synth_rep} \\
{\bf h}^R_r &= \texttt{RGroupEncoder}(\{{\bf h}^S_s: \forall s\in \sigma(r)\}), \label{group_rep} \\
{\bf h}_t^T &= \texttt{ReactionEncoder}(\{{\bf h}^R_{r}: \forall r\in \psi(t)\}). \label{rxn_rep} \end{align}

In \eqref{likelihood}, we considered a factorization of the likelihood such that, given a molecular representation ${\bf z}$, the molecular decoder proceeds by first decoding the reaction type and then decoding one synthon for each R-group separately conditional on the reaction type. Hence, we require a key vector for each reaction as well as for each synthon to compare with the associated query vectors (to be described). As such, we introduce a $\texttt{ReactionKeyGenerator}: \mathbb{R}^{d_T}\to\mathbb{R}^{k_T}$ which returns a key vector given a reaction representation, and similarly introduce a $\texttt{SynthonKeyGenerator}: \mathbb{R}^{d_S}\to\mathbb{R}^{k_S}$ to produce a key for each synthon. Each of these key generators can be parameterized by MLPs:
\begin{align}
{\bf k}_t^T &= \texttt{ReactionKeyGenerator}({\bf h}_t^T), \label{rxn_key} \\
{\bf k}_s^S &= \texttt{SynthonKeyGenerator}({\bf h}_s^S). \label{synth_key}
\end{align}

\subsubsection{Molecular encoder}

The $\texttt{MolecularEncoder}:\mathscr{G}_X\to\mathbb{R}^{d_X}$ takes as input a molecular graph $\mathcal{G}_x\in\mathscr{G}_X$ and returns a $d_X$-dimensional feature representation,
\begin{align}
{\bf z} &= \texttt{MolecularEncoder}(\mathcal{G}_x). \label{mol_query}
\end{align}

In our implementation, the \texttt{MolecularEncoder} is a graph neural network with a variational linear layer stacked on top of the readout, which produces a sample ${\bf z}\sim q({\bf z}|x)$ for a given input graph $\mathcal{G}_x$. As depicted in Figure \ref{figure1}, we interpret the representation ${\bf z}$ as a query induced by $x$ into the library $\mathcal{D}$.

Note that the \texttt{MolecularEncoder} can in principle take any valid molecular graph as input is therefore capable of producing queries for compounds that are \emph{not} in the library $\mathcal{D}$. This can be useful for finding \emph{analogues by catalog} -- that is, compounds which can be purchased from a catalog and are chemical analogues of a query molecule. This use case will be investigated in a later section.

\subsubsection{Molecular decoder}

Given a query ${\bf z}=\texttt{MolecularEncoder}(\mathcal{G}_x)$ and a library $\mathcal{D}$, the decoder is tasked with retrieving the molecule from the library, i.e., identifying the reaction and synthon tuple which yield the molecule as a product. We proceed by generating the reaction first. A $\texttt{ReactionQueryGenerator}: \mathbb{R}^{d_X}\to \mathbb{R}^{k_T}$ generates a query from the molecular representation to compare against the reaction keys:
\begin{align}
{\bf q}^T &= \texttt{ReactionQueryGenerator}({\bf z}), \label{rxn_query} \\
\Big\{p(t_j|{\bf z},\mathcal{D})\Big\}_{j=1}^{|T|} &= \text{softmax}\left(\left\{\frac{{\bf q}^T \cdot {\bf k}_{t_j}^T}{\sqrt{k_T}}\right\}_{j=1}^{|T|}\right). \label{rxn_probs}
\end{align}

This defines a probability distribution over reaction types in $T$. We can sample according to this distribution to arrive at a reaction $t\sim p(t|{\bf z},\mathcal{D})$.

Given the sampled reaction $t$, we know the required R-groups (via $\psi$) and further know which synthons are eligible for each R-group (via $\sigma$). To decode the synthon tuple, we introduce a $\texttt{SynthonQueryGenerator}: \mathbb{R}^{d_X+d_T+d_R}\to\mathbb{R}^{k_S}$ which can be used to query synthons for each R-group $r_t^{(i)}\in \psi(t)$ as follows:
\begin{align}
{\bf q}^S_{t,r} &= \texttt{SynthonQueryGenerator}([{\bf z} \| {\bf h}_t^T \| {\bf h}_r^R]), \label{synth_query} \\
\left\{p(s_j^{(i)}|{\bf z},t, r^{(i)}_t,\mathcal{D})\right\}_{j=1}^{|\sigma(r^{(i)}_t)|} &= \text{softmax}\left(\left\{\frac{{\bf q}_{t,r^{(i)}_t}^S\cdot {\bf k}^S_{s_j^{(i)}}}{\sqrt{k_S}}\right\}_{j=1}^{|\sigma(r_t^{(i)})|}\right). \label{synth_probs}
\end{align}

In our implementation, both the reaction and synthon query generators are parameterized by MLPs.

\subsection{Training algorithm}

For a large CSL $\mathcal{D}^*$, encoding the entire library on each iteration of the training loop could require an excessive amount of GPU memory. To overcome this, we utilize a minibatch strategy in which a random subset $\mathcal{D}\subset\mathcal{D}^*$ is drawn from the full library according to a distribution $p({\mathcal{D}}|\mathcal{D}^*)$. From $\mathcal{D}$, we form the synthon, R-group, and reaction representations and keys. In particular, we use a sub-sampler $p({\mathcal{D}}|\mathcal{D}^*)$ which (i) samples a subset of the reactions contained in the full library uniformly at random, keeping only the R-groups contained in the sampled reactions, and (ii) for each reaction, samples a random number of products, retaining only the synthons that are contained in the sampled products. We also utilize teacher forcing, feeding in the ground truth reaction when generating the synthon queries for the respective R-groups. Algorithm \ref{alg:training} in the \hyperref[sec:appendix]{Appendix} describes the training procedure.

\paragraph{Ex-post density estimation} 

Given a trained generative model $p_\theta(x|{\bf z})$, we wish to sample products via $(x,{\bf z})\sim p_\theta(x|{\bf z})\,p({\bf z})$ (discarding ${\bf z}$).\footnote{The parameter $\theta$ represents the parameters for the modules written in \texttt{typewriter} font.} However, this in general will not correspond well to a uniform distribution over the products in $\mathcal{D}$ due to the bias introduced by the batch sampling strategy outlined above (which first uniformly samples reactions and then uniformly samples products given the reaction).

Although this can be corrected with importance weighting during the training phase, we opt for a more practical approach by using an ex-post density estimation strategy \cite{ghosh2019variational}. We sample a large number of products from the target distribution $x\sim p(x|\mathcal{D})$ and encode these products via the molecular encoder ${\bf z}\sim q_\phi({\bf z}|x)$. We then fit a density estimator to the aggregated samples, written $q_\lambda({\bf z})$. In our experiments, we utilize a multivariate normal distribution for simplicity, but one could imagine using more expressive density estimators here (e.g., a mixture of multivariate normals).

Now, we can sample products via $(x, {\bf z})\sim p_\theta(x|{\bf z})\,q_\lambda({\bf z})$, which will more closely align with sampling from $p(x|\mathcal{D})$. This helps to correct for bias in the distribution over product space that is induced by the choice of batch sampling strategy.

\subsection{Computational complexity, scalability, and efficiency}

We now examine the computational complexity, scalability, and efficiency of the proposed method.

First, we note that the CSL $\mathcal{D}$ can be encoded with $O(|S| + |R| + |T|)$ complexity -- the constant depends on the complexity of the synthon, R-group, and reaction encoders. Nonetheless, this is logarithmic in comparison to naively encoding each product in $\mathcal{D}$, which has $O(|\mathcal{D}|)$ complexity. 

More noteworthy is the computational complexity of the molecular decoder. For clarity, let us consider a simplified $\mathcal{D}$ comprised of a single $k$-component reaction. Let $M_i$ denote the number of synthons for R-group $i=1,\ldots,k$. Naively, a nearest neighbor lookup in $\mathcal{D}$ requires $O(\prod_{i=1}^k M_i)$ complexity. CSLVAE, on the other hand, performs the lookup over synthons in each R-group independently, which attains $O(\sum_{i=1}^k M_i)$ complexity: a logarithmic improvement. Hence, the proposed molecular decoder is highly suitable for ultra-large CSLs that are of interest in early-stage drug discovery.

Another advantage of CSLVAE's decoding strategy is that it relies only minimally on autoregression. In fact, we only ever need to do a single step of autoregression, irrespective of the size of the graph being generated (autoregression length of exactly two). As such, CSLVAE gracefully scales to large and variable-sized molecular graphs that follow a combinatorial synthesis construction.

Lastly, we point out that our method is guaranteed to generate chemically valid---and perhaps more importantly, synthetically accessible---molecular graphs without performing explicit validity checks. This compares favorably with prior work, in which the validity of each candidate action is verified at each step of the autoregression, with invalid actions excluded from the choice set. Although cheminformatics libraries like RDKit \cite{rdkit} have efficient C++ implementations for these checks, they nonetheless increase runtime rather significantly. Further, in the absence of explicit validity checks, these models have been shown to generate invalid molecular graphs at a markedly higher rate \cite{jin2018junction}.

\section{Experiments}

This section covers some of our attempts to validate CSLVAE's performance and highlight its capabilities. We include additional supplementary experiments in the \hyperref[sec:appendix]{Appendix}.

\paragraph{Data} We demonstrate the capabilities of CSLVAE on the Enamine REAL library, which is comprised of 340K synthons and over one thousand reactions. The reactions in REAL range from two to four components and the number of synthons per R-group range from the single digits to tens of thousands. In total, the REAL library describes a chemical space of over 16 billion commercially available compounds.\footnote{We release a subset of the library alongside our code for reproducing these experiments and to foster further research in the machine learning community applied to combinatorial synthesis libraries: \href{https://github.com/AtomwiseInc/cslvae}{\texttt{https://github.com/AtomwiseInc/cslvae}}. Data provided with permission from and attribution to Enamine Ltd.} Note that this 16 billion compound library is relatively small compared to over-trillion compound libraries that are commercially available today; we use this more modest library size as it makes comparisons to other approaches tractable.

\paragraph{Training} During training, we sample subsets of the library as follows. Of the roughly 1300 reaction types in the REAL database, we first uniformly sample 20 reactions at random, and subsequently sample 100 products per reaction, including the associated synthons in the library subset. These library subsets therefore describe roughly 300K-1.5M compounds each, which is significantly smaller than the complete library of 16 billion compounds. See Algorithm \ref{alg:training} in the \hyperref[sec:appendix]{Appendix} for details.

\paragraph{Testing} For test-time inference, we decode with respect to the full library of 16 billion compounds. This constitutes a test-time distribution shift relative to training, but we observed that CSLVAE generalizes remarkably well to the full library without modifications. For completeness, we include an analysis of the test-time distribution shift in the \hyperref[sec:appendix]{Appendix} (Supplementary Figure \ref{fig:testshift}). In the forthcoming analyses, we share results when performing inference on the full library, as this is our primary objective. 

\subsection{Molecular reconstruction and generation}

\begin{table}[h]
\centering
\caption{Comparison of RationaleRL, JT-VAE, and CSLVAE (ours) on synthon-based generative modeling.}
\begin{tabular}{l| c c c}
\hline
 & JT-VAE & RationaleRL & CSLVAE (ours) \\
\hline
\# Parameters & 4.7M & 3.4M & 380K \\
\hline
Validity & 100.0\% & 100.0\% & 100.0\% \\
Uniqueness & 80.1\% & 96.3\% & 98.8\% \\
Average likelihood & 18.7\% & 62.3\% & 72.4\% \\
\hline
In-library proportion & 2.9\% & 50.9\% & 100.0\% \\
\hline
\end{tabular}
\label{tab:comparison}
\end{table}

We compare CSLVAE against two state-of-the-art molecular graph generative models: JT-VAE \cite{jin2018junction} and RationaleRL \cite{jin2020multi}. All three models were trained from scratch on the Enamine REAL library. Details on the experimental setup and architecture can be found in the \hyperref[sec:appendix]{Appendix}.

In JT-VAE, molecular graphs are represented by junction trees over chemical fragments. Decoding proceeds by first generating the junction tree in a depth-first manner, placing a fragment in each node, and then subsequently orienting the fragments to match attachment points. RationaleRL, on the other hand, takes as input a starting chemical fragment or {\em rationale}. The decoder's objective is to complete the molecule in an autoregressive fashion (one graph edit per step). In our setting, we take a product from the library and remove all but one synthon, treating the resulting graph as the starting rationale. Thus, RationaleRL is tasked with generating the missing synthon.

Table \ref{tab:comparison} summarizes the key findings. First, we note that our implementation of CSLVAE has roughly 10x fewer parameters than the two alternatives considered, owing to the inductive nature of the library encoder. All three methods achieve 100\% chemical validity, but CSLVAE achieves this result without explicit validity checks. The average likelihood is computed by taking the average of per-compound reconstruction likelihoods across a large number of products sampled from the library. This is a measure of how well the model is capable of reconstructing the full molecular graph  (i.e., on average, how likely are we to reproduce the query molecule via the decoder) and can also loosely be interpreted as a measure of coverage/reachability (i.e., what proportion of the library are we able to faithfully cover). Finally, we highlight the challenges existing graph generative models face when applied to ultra-large CSLs, namely that they struggle to reliably generate in-library compounds. For JT-VAE, fewer than 1 in 34 compounds were found in REAL. RationaleRL, on the other hand, generates in-library completions in only about half of the cases (see Supplementary Figure \ref{fig:extrapolation} in the \hyperref[sec:appendix]{Appendix}), but has the advantage that it is provided a starting rationale in the form of a compound from REAL stripped of all but one synthon. In contrast, CSLVAE is guaranteed to stay in the library by design. 

\subsection{Latent space visualizations}

Next, we qualitatively inspect the latent space learned by CSLVAE. In particular, we are interested in verifying whether the proposed model has learned a latent space which varies relatively smoothly over the covered chemical space (i.e., that small perturbations to the query induce only minor edits in the resulting molecular graph). We perform two kinds of checks: latent space interpolations and local neighborhood visualizations.

\begin{figure}
    \centering
    \includegraphics[origin=c,width=5in]{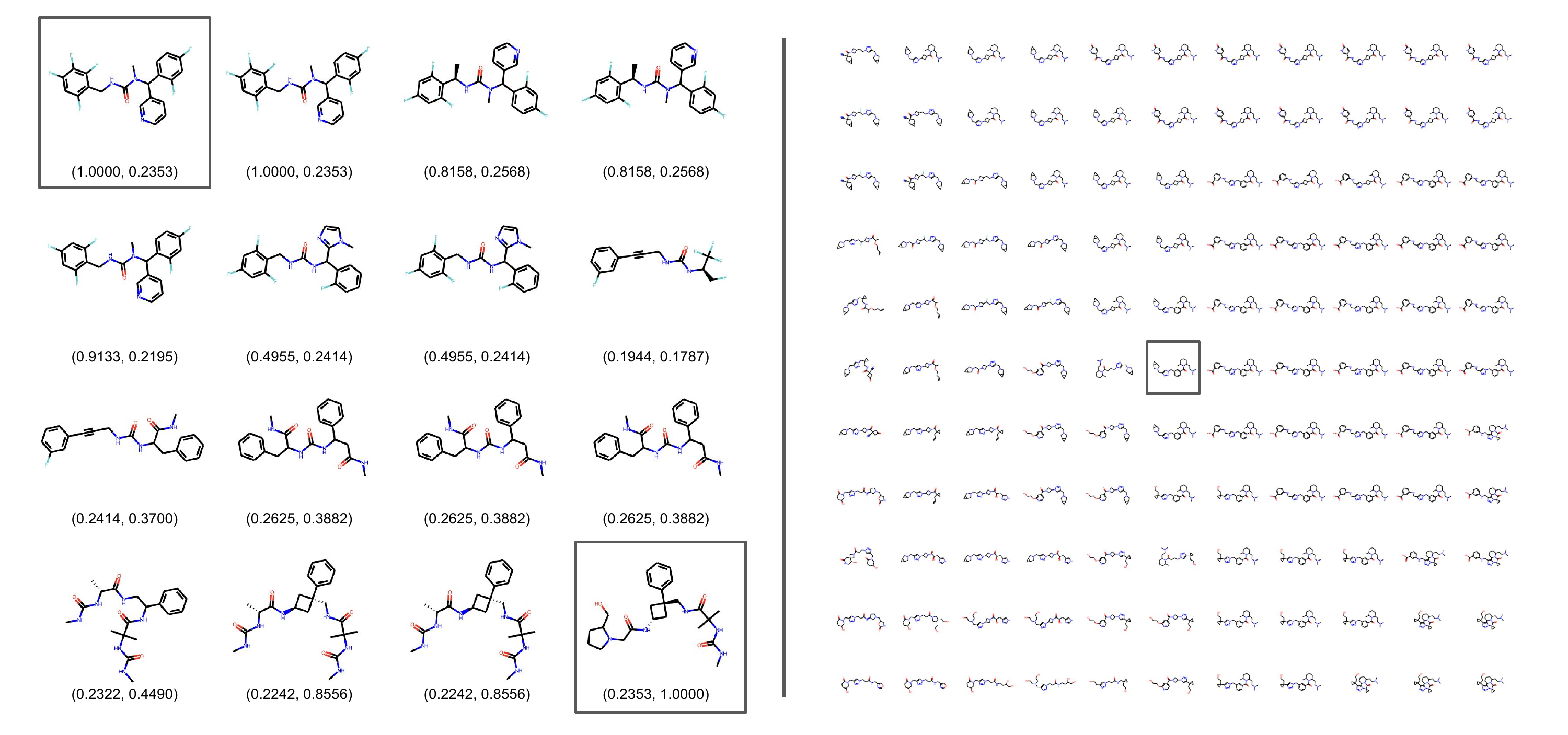}
    \caption{\textbf{Latent space visualizations.} \textbf{Panel A:} Moving left to right in raster order, we linearly interpolate from the starting compound to the target compound. The immediately adjacent molecules are reconstructions. Below each molecule are its Tanimoto similarities with the starting and target compound, respectively. \textbf{Panel B:} We sample two random directions in the latent space around a query compound and visualize the decoded molecules spaced evenly on the resulting 2D plane.}
    \label{latentspaceviz}
\end{figure}

Panel A of Figure \ref{latentspaceviz} contains an example of interpolations in the latent space. In particular, we interpolate the molecular queries for the starting compound (top left) and target compound (bottom right), in raster order. The molecules immediately adjacent to the starting and target compounds are the associated reconstructions. Products are decoded with respect to the full REAL library of 16 billion compounds. Below each molecule are its Tanimoto similarity\footnote{The Tanimoto similarity \cite{bajusz2015tanimoto} is calculated by taking the intersection-over-union between a pair of bit vectors describing each molecule using a hash called a \emph{molecular fingerprint} \cite{bender2004molecular}.} with the starting and target molecule, respectively. We observe that the interpolations traverse through regions of chemical space that gradually decrease (cf. increase) in similarity with respect to the starting (cf. target) compound.

Panel B of Figure \ref{latentspaceviz} visualizes the latent space around a randomly sampled product from the REAL library. Following prior work \cite{kusner2017grammar}, we form a random 2D plane in the high-dimensional latent space by sampling two random directions around the molecular query (center compound) and decoding the resulting products using the argmax decision rule. We observe that the latent space is smooth in the sense that molecules morph gradually, with only minor edits when the movement in latent space is small (e.g., one synthon at a time, modifications to smaller functional groups), and that the molecular scaffold is generally conserved locally.

\subsection{Analogue retrieval via autoencoding}

\begin{figure}
    \centering
    \includegraphics[origin=c,width=5in]{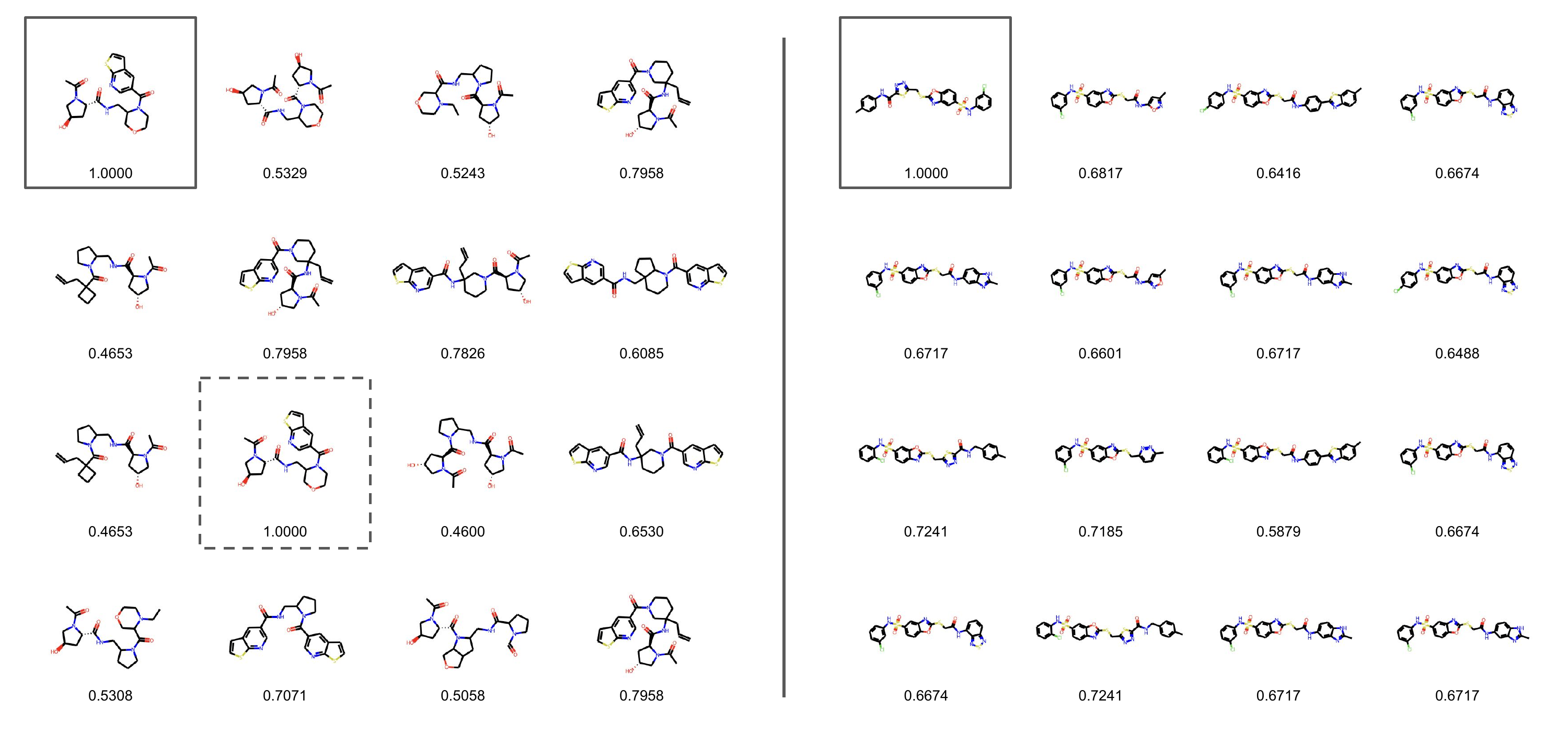}
    \caption{\textbf{Analogue retrieval via autoencoding.} Each query molecule (top left) is encoded and stochastically decoded fifteen times. Below each molecule is its Tanimoto similarity with the query molecule. \textbf{Left:} Autoencoding a molecule contained in the library. \textbf{Right:} Autoencoding a molecule not contained in the library.}
    \label{analogues}
\end{figure}

Lastly, we utilize CSLVAE to find analogues of a query compound in a large CSL. In Figure \ref{analogues}, we present the model with two molecules: one which is in the library (left) and one which is not in the library (right). Given the molecular query, we generate conditionally random completions from the decoder. In both cases, autoencoding retrieved highly similar compounds as measured by Tanimoto similarity. For the in-library example, we note that the model is able to successfully retrieve the query compound (third row, second column). Moreover, we find many instances in which autoencoding returns compounds that have the same reaction type as the query molecule, but vary in one or two synthons. For the out-of-library example, we note that CSLVAE finds skeletally relevant compounds with high Tanimoto similarity, indicating the promise of this approach for fast analogue search in large (non-enumerable) libraries.

\section{Closing remarks}

We proposed the combinatorial synthesis library variational auto-encoder, or CSLVAE, a new graph-based generative model for the navigation of combinatorial synthesis libraries. CSLVAE utilizes minimal autoregression, permitting efficient generation of large molecular graphs and improving scalability. Compounds generated by CSLVAE are chemically valid and cost-effectively accessible. CSLVAE is a neural database providing random access to non-enumerable libraries. In experiments, we demonstrate the capabilities of CSLVAE in modeling ultra-large and realistic make-on-demand libraries, paving a path towards more scalable strategies in the exploration of non-enumerable chemical libraries for early-stage drug discovery. Our method can be combined with established techniques for molecular optimization as a future direction.

Our approach has some limitations; here, we highlight three. First, the synthon lookup in the decoder scales linearly with the number of synthons in an R-group, which can present challenges as libraries continue to add many synthons per R-group. This could be mitigated with more scalable query-key designs \cite{choromanski2020rethinking,kitaev2020reformer}, but is not considered here. Furthermore, CSLVAE may face difficulties in the presence of prominent R-group symmetry (e.g., as in polymers). The decoder would require modifications to break parity, but may not admit the same convenient parallelization. Lastly, softmax has limitations in mapping from real-valued potentials to choice probabilities due to its rigid substitution patterns \cite{train2009discrete}; sparse or alternative-aware softmax variants may be preferable, but are not considered in this paper. In future work, we intend to address these shortcomings and look to applications in virtual high throughput screening.

\begin{ack}
This work is sponsored by Atomwise Inc. The authors would like to give special acknowledgement to Christian Laggner, Ho Leung Ng, Srimukh Prasad, Adrian Stecula, and Brad Worley for discussions and helpful suggestions.
\end{ack}

\bibliography{main}

\newpage

\appendix

\renewcommand{\figurename}{Supplementary Figure}
\renewcommand{\tablename}{Supplementary Table}

\setcounter{figure}{0}
\setcounter{table}{0}

\label{sec:appendix}

\section{CSLVAE details}

This section provides additional details on CSLVAE omitted from the main paper due to the page limit.

\subsection{Training and ex-post density estimation algorithms}

We describe the CSLVAE training procedure in Algorithm \ref{alg:training} and the CSLVAE ex-post density estimation procedure in Algorithm \ref{alg:expost}.

\begin{algorithm}
\caption{CSLVAE training procedure}\label{alg:training}
\begin{algorithmic}
\State {\bf input} Full library $\mathcal{D}^*$, library sub-sampler $p(\mathcal{D}œ|\mathcal{D}^*)$, batch size $N$, model parameters $\theta$, KL divergence weight $\beta\ge 0$, choice of optimizer
\While{\emph{stopping criteria not reached}} \\
\quad\,\emph{\# Prepare a minibatch} \\
\quad\, Sample a library $\mathcal{D}\sim p(\mathcal{D}|\mathcal{D}^*)$ \\
\quad\, Sample a minibatch of reaction-synthon chain pairs, $(t_n, u_n) \sim p(t,u|\mathcal{D})$ for $n=1\ldots N$ \\
\quad\, Form the corresponding products via synthesis, $x_n=f(t_n,u_n)$ for $n=1 \ldots N$ \\
\quad\, \emph{\# Encode the library} \\
\quad\, Get the synthon representations, $\mathbf{h}_s^S$ for each $s\in S$, as per equation (5) \\
\quad\, Get the R-group representations, $\mathbf{h}_r^R$ for each $r\in R$, as per equation (6) \\
\quad\, Get the reaction representations, $\mathbf{h}_t^T$ for each $t\in T$, as per equation (7) \\
\quad\, Get the reaction keys, ${\bf k}_t^T$ for each $t\in T$, as per equation (8) \\
\quad\, Get the synthon keys, ${\bf k}_s^S$ for each $s\in S$, as per equation (9) \\
\quad\, \emph{\# Encode the products} \\
\quad\, Sample the product queries, ${\bf z}_n$ for each $n=1 \ldots N$, as per equation (10) \\
\quad\, Calculate the KL divergence contributions, $\text{KLD}_n = \text{D}_\text{KL}(q({\bf z}|x_n)\| p({\bf z}))$ for each $n=1\ldots N$\\
\quad\, \emph{\# Decode the products wrt the library} \\
\quad\, Get the reaction queries, $[\mathbf{q}^T]_n$ for $n=1\ldots N$, as per equation (11) \\
\quad\, Get the reaction probabilities, $p(t|{\bf z}_n,\mathcal{D})$ for each $t\in T$ and $n=1\ldots N$, as per equation (12) \\
\quad\, Get the synthon queries, $[\mathbf{q}^S_{t_n,r}]_n$ for each $n=1\ldots N$ and $r\in \sigma(t_n)$, as per equation (13) \\
\quad\, Get the synthon probabilities, $p(s_j^{(i)}|{\bf z}_n,t_n,r_{t_n}^{(i)},\mathcal{D})$ for each $n=1\ldots N$, $s^{(i)}_j\in \psi(r_{t_n}^{(i)})$, and \\
\quad\quad\, $r_{t_n}^{(i)}\in \sigma(t_n)$, as per equation (14) \\
\quad\, Calculate the LL contributions, $\text{LL}_n = \log p(t_n|{\bf z}_n,\mathcal{D}) + \sum_{i=1}^{|u_n|}\log p(u_n^{(i)}|{\bf z}_n,t_n,r_{t_n}^{(i)},\mathcal{D})$ for \\
\quad\quad\, each $n=1\ldots N$ \\
\quad\, \emph{\# Compute the loss and update parameters} \\
\quad\, Calculate the loss, which is the negative of the weighted ELBO, $\ell=\frac{1}{N}\sum_{n=1}^N \beta\text{KLD}_n-\text{LL}_n$ \\
\quad\, Compute gradients $\nabla_\theta \ell$ and update $\theta$ using choice of optimizer, i.e., $\theta\leftarrow\text{optimizer}(\theta,\nabla_\theta\ell)$
\EndWhile \\
\textbf{return} Fitted model parameters $\theta^*\leftarrow\theta$
\end{algorithmic}
\end{algorithm}

\begin{algorithm}
\caption{CSLVAE ex-post density estimation procedure}\label{alg:expost}
\begin{algorithmic}
\State {\bf input} Master library $\mathcal{D}$, number of examples $N$, fitted model parameters $\theta^*$, density estimator $q_\lambda$
\For{$n=1,\ldots,N$} \\
\quad\, Sample $x_n\sim p(x|\mathcal{D})$, as per equation (1) \\
\quad\, Sample a product query ${\bf z}_n\sim q_{\theta^*}({\bf z}|x_n)$ as per equation (10) 
\EndFor \\
Fit density $q_\lambda({\bf z})$ to the aggregated product queries $\{{\bf z}_n\}_{n=1}^N$ \\
\textbf{return} Fitted density estimator $q_{\lambda^*}\leftarrow q_{\lambda}$
\end{algorithmic}
\end{algorithm}

\subsection{Architecture}

Below, we walk through the CSLVAE architecture utilized in our experiments. Supplementary Table \ref{tab:cslvae_arch} summarizes the overall CSLVAE architecture used in our experiments. In total, the version of CSLVAE we utilized is comprised of 384,512 learnable parameters.

\subsubsection{Atom and bond embeddings}

Both the synthon encoder and molecular encoder, which are parameterized as message passing graph neural networks, take shared atom and bond embeddings as the initial node and edge features, respectively. Atoms and bonds are represented by a set of binary features, which we described in Supplementary Tables \ref{tab:atomembedder} and \ref{tab:bondembedder}, respectively. In our implementation, the node and edge dimensions are set to 64 each. As such, the atom embeddings consist of 50 $\times$ 64 = 3,200 parameters, and the bond embeddings consist of 12 $\times$ 64 = 768 parameters.

\begin{table}[ht]
\centering
\caption{Atom feature types.}
\begin{tabular}{c | c | c }
\hline
Feature & Choices & Number of choices \\
\hline
Element type & *, B, Br, C, Cl, F, Fe, I, N, O, P, S, Se, Si, Sn & 15 \\
Node degree & 1, 2, 3, 4, 5, 6, $\ge$7 & 7 \\
Hybridization & S, SP, SP2, SP3, SP3D, SP3D2, Unspecified & 7 \\
Chirality & CW, CCW, Other, Unspecified & 4 \\
Bonded hydrogens & 0, 1, 2, 3, 4, 5, 6, $\ge$7 & 8 \\
Formal charge & $\le$-3, -2, -1, 0, 1, 2, $\ge$3 & 7 \\
Aromatic & False, True & 2 \\
\hline
& & 50 \\
\hline
\end{tabular}
\label{tab:atomembedder}
\end{table}

\begin{table}[ht]
\centering
\caption{Bond feature types.}
\begin{tabular}{c | c | c }
\hline
Feature & Choices & Number of choices \\
\hline
Bond type & Single, Double, Triple, Aromatic & 4 \\
Conjugated & False, True & 2 \\
In a ring & False, True & 2 \\
Stereochemistry & One, Z, E, Any & 4 \\
\hline
& & 12 \\
\hline
\end{tabular}
\label{tab:bondembedder}
\end{table}

\subsubsection{Synthon encoder}

The \texttt{SynthonEncoder} is a message passing graph neural network, taking node, edge, and graph features as input, and producing updates to these features after every round of message passing in a residual fashion. We describe a single message passing layer below.

\paragraph{Preliminaries} A graph $\mathcal{G}=(\mathcal{V},\mathcal{E})$ is represented by a set of node features $\{{\bf x}_i^0 : i\in\mathcal{V}\}$, edge features $\{{\bf e}_{ij}^0: (i, j)\in\mathcal{E}\}$, and graph features ${\bf g}^0$. Superscripting by zero indicates that these are the initial or input features; below, superscripting by $\ell$ denotes the result after the $\ell$th round of message passing.

\paragraph{Edge model} The edge model updates the edge features as follows. For an edge $(i,j)\in\mathcal{E}$, we concatenate the current iterate $\ell$ of the edge features, graph features, and the two corresponding node features. The concatenated features are then layer normalized and processed by a two-layer MLP, producing $\Delta{\bf e}^{\ell+1}_{ij}=\texttt{Linear}(\texttt{ReLU}(\texttt{Linear}(\texttt{ReLU}(\texttt{LayerNorm}([{\bf e}_{ij}^\ell\|{\bf g}^\ell\|{\bf x}_i^\ell\|{\bf x}^\ell_i])))))$. The output dimension of the final linear layer is set to be equal to the dimensionality of the edge features, which allows the edge features to be updated in a residual fashion, i.e., ${\bf e}_{ij}^{\ell+1}={\bf e}_{ij}^\ell + \Delta{\bf e}_{ij}^{\ell+1}$. Note that for an edge $(i,j)\in\mathcal{E}$, we separately maintain edge features for $i\to j$ and $i\leftarrow j$ updates; these features are identical for $\ell=0$, but are not the same in general for subsequent layers.

\paragraph{Node model} Given a node $i\in\mathcal{V}$, we first form a message from all its neighboring nodes by summing over the associated incoming edge features, ${\bf m}_i^{\ell+1}=\sum_{j\in\mathcal{N}(i)}{\bf e}^{\ell+1}_{ij}$. Similar to the edge model, the node feature, message, and graph feature are concatenated, layer normalized, and passed through a two-layer MLP in which the final linear layer has output dimension equal to the dimensionality of the node features. This produces a residual update for the node feature, given by $\Delta{\bf x}^{\ell+1}_{i}=\texttt{Linear}(\texttt{ReLU}(\texttt{Linear}(\texttt{ReLU}(\texttt{LayerNorm}([{\bf x}_{i}^\ell\|{\bf m}^{\ell+1}_i\|{\bf g}^\ell])))))$. The node features are thus updated according to ${\bf x}^{\ell+1}_i={\bf x}^\ell_i+\Delta{\bf x}^{\ell+1}_i$.

\paragraph{Graph model} The graph features are updated by accumulating messages from each node in $\mathcal{V}$ as follows. Node messages to the graph $\{{\bf n}_i^{\ell+1}: i\in\mathcal{V}\}$ are formed using the now familiar layer norm + MLP design, with the final linear layer having output dimension equal to the dimensionality of the graph feature. These messages, ${\bf n}^{\ell+1}_{i}=\texttt{Linear}(\texttt{ReLU}(\texttt{Linear}(\texttt{ReLU}(\texttt{LayerNorm}([{\bf x}_{i}^\ell\|{\bf m}^{\ell+1}_i\|{\bf g}^\ell])))))$, are then aggregated via sum pooling to form a residual to the graph features, $\Delta{\bf g}^{\ell+1}=\sum_{i\in\mathcal{V}}{\bf n}^{\ell+1}_i$, which updates the graph features accordingly, ${\bf g}^{\ell+1}={\bf g}^\ell+\Delta{\bf g}^{\ell+1}$. Implementation-wise, the node and graph model use a shared MLP with the output dimension equal to the sum of the node and graph feature dimensions, and the outputs are then split into two parts: one which routes to the node update, and the other to the graph update.

\paragraph{Putting it all together} The message passing neural network applies a sequence of message passing layers as outlined above. In our implementation, we set the node, edge, and graph feature dimensions to 64 and utilize four message passing layers. For the input graph features, we simply use a vector of zeros. The final graph features serve as the synthon representations, which are used in producing synthon keys and the cascaded representations for the R-groups and reactions. In total, the \texttt{SynthonEncoder} we utilize in experiments is described by 152,832 parameters.

\subsubsection{R-group encoder}

The \texttt{RGroupEncoder} follows the design outlined in DeepSets, wherein the synthon representations are each processed separately by an MLP, pooled together (here, we use mean pooling to allow the network to focus on characteristics of the \emph{distribution} of synthons belonging to an R-group), and the result processed by yet another MLP. In our experiments, both MLPs are set as two-layer networks with ReLU activation in between the two linear operations. All dimensions are 64, and as such the \texttt{RGroupEncoder} utilizes 16,640 parameters in total.

\subsubsection{Reaction encoder}

The \texttt{ReactionEncoder} follows the same design as the \texttt{RGroupEncoder}, with the exception that sum pooling is utilized instead of mean pooling to allow the network to focus on the multi-set of R-groups in a reaction. All dimensions are 64, and as such the \texttt{ReactionEncoder} also utilizes 16,640 parameters in total.

\subsubsection{Synthon key generator}

The \texttt{SynthonKeyGenerator} is an MLP which produces a synthon key from a synthon representation. In our implementation, we simply use a linear layer. The input and output dimensions are both set to 64, and as such the \texttt{SynthonKeyGenerator} utilizes 4,160 parameters in total.

\subsubsection{Reaction key generator}

The \texttt{ReactionKeyGenerator} is an MLP which produces a reaction key from a reaction representation. In our implementation, we simply use a linear layer. The input and output dimensions are both set to 64, and as such the \texttt{ReactionKeyGenerator} utilizes 4,160 parameters in total.

\subsubsection{Molecular encoder}

The \texttt{MolecularEncoder} produces molecular queries and utilizes the same message passing design as the \texttt{SynthonEncoder}, so refer to the early subsection describing the architecture. Again, the node, edge, and graph features are all set to 64, and we use four layers of message passing. However, differently from the \texttt{SynthonEncoder}, the \texttt{MolecularEncoder} has an additional variational linear layer that produces a conditional mean and conditional log variance vector from the graph features produced by the final message passing round. The variational linear layer takes a 64 dimensional graph feature as input and produces a 128 dimensional output, which is split into the mean and log variance portions. As such, the \texttt{MolecularEncoder} utilizes a total of 152,832 + 8,320 = 161,152 parameters.

\subsection{Molecular query processing network}

The molecular queries are regularized to be close to the prior $p({\bf z})$ as a consequence of the VAE objective. To allow the decoder to make better use of these features, the molecular queries are processed by an MLP prior to being passed to the reaction and synthon query generators. Opting for simplicity, we use a simple two-layer MLP with intermediate ReLU activation; all dimensions are 64, which results in 8,320 parameters.

\subsubsection{Reaction query generator}

The \texttt{ReactionQueryGenerator} is an MLP which produces a reaction query from the molecular query. In our experiments, we utilize a two-layer network with intermediate ReLU activation. All dimensions are set to 64, so the \texttt{ReactionQueryGenerator} utilizes 8,320 parameters in total.

\subsubsection{Synthon query generator}

Finally, the \texttt{SynthonQueryGenerator} is an MLP with produces a synthon query from the molecular query, reaction representation, and R-group representation. In the general case, we suggest concatenating these three feature types; because we have utilized a common dimensionality of 64 throughout and to keep implementation simple, we use sum instead of concatenation in our experiments (which can be shown to be equivalent to concatenating with an additional constraint on the weight matrix for the subsequent linear layer). Like the \texttt{ReactionQueryGenerator}, we also use a two-layer MLP with intermediate ReLU activation, resulting in a total of 8,320 parameters for the \texttt{SynthonQueryGenerator}.

\begin{table}[ht]
\centering
\caption{Summary of CSLVAE architecture used in experiments.}
\begin{tabular}{c | c | c }
\hline
Module & Module type & Number of parameters \\
\hline
Atom embedding & Embedding & 3,200 \\
Bond embedding & Embedding & 768 \\
Synthon encoder & GNN & 152,832 \\
R-group encoder & DeepSets & 16,640 \\
Reaction encoder & DeepSets & 16,640 \\
Synthon key generator & Linear & 4,160 \\
Reaction key generator & Linear & 4,160 \\
Molecular encoder & GNN & 161,152 \\
Molecular query processing network & MLP & 8,320 \\
Reaction query generator & MLP & 8,320 \\
Synthon query generator & MLP & 8,320 \\
\hline
& & 384,512 \\
\hline
\end{tabular}
\label{tab:cslvae_arch}
\end{table}

\section{Comparison of CSLVAE to RationaleRL and JT-VAE}

In this section, we provide additional details on the experiment comparing CSLVAE with two state-of-the-art graph generative models, RationaleRL and JT-VAE, which is summarized by Table \ref{tab:comparison} in the body of the paper.

\subsection{Dataset preparation}
The Enamine REAL library is a combinatorial synthesis library of roughly 1300 reaction types, ranging from 2- to 4-component reactions, along with roughly 340K synthons. In total, the REAL library describes a chemical space in excess of 16B make-on-demand compounds. For the two baseline models we compare against, we faced memory issues when using the author-provided code on the full REAL library (both in the vocabulary generation steps as well as with writing products to disk). As such, the two baselines were trained and evaluated on a subset of the full REAL library. We did not face such memory limitations with CSLVAE, which we trained on the full library and evaluated as such. Hence, RationaleRL and JT-VAE can be compared on a per-item basis (each having been trained on the same subset of REAL), whereas our results for CSLVAE reflect training on the larger and more diverse full REAL library. We note that RationaleRL and JT-VAE were not developed with the goal of searching through combinatorial synthesis libraries in mind, and our use of them serves as an attempt to compare our method with the application of existing state-of-the-art graph generative model on combinatorial synthesis libraries out-of-the-box and without modification.

To construct the data on which RationaleRL and JT-VAE were trained, we ranked the $\sim$1300 reaction types by the number of products contained in each reaction and selected 50 of the middle-sized reactions. In total, these reactions describe a chemical space of $\sim$125M compounds. We sampled products from each reaction such that all synthons are represented, amounting to a training set of $\sim$500K compounds.

\subsection{RationaleRL details}\label{rrl}
We utilize the \emph{pre-training} phase of RationaleRL, which trains a graph-based variational autoencoder that seeks to reconstruct a molecular graph from a \emph{starting rationale} (and does not require RL, as the name might suggest; that is part of the \emph{fine-tuning} phase, which we do not utilize here). Given the starting rationale and full molecule, RationaleRL's decoder completes the molecule autoregressively in an atom-by-atom, bond-by-bond fashion. To form the starting rationale, we take a product from REAL and remove all but one synthon. Hence, RationaleRL is tasked with completing the missing synthon given the full molecule (as input to the encoder) and the starting rationale (as input to the decoder, along with the latent code). In Supplementary Figure \ref{fig:extrapolation}, we show that as the number of distinct reactions in training is increased, the probability that completions generated by RationaleRL yield a product from the library drops precipitously; after 50 reactions are included, fewer than half of all completions yield a product in the REAL library. Our experiments are based on the author-provided code, which can be found at \href{https://github.com/wengong-jin/multiobj-rationale}{\texttt{https://github.com/wengong-jin/multiobj-rationale}}.

\begin{figure}
    \centering
    \includegraphics[width=0.8\linewidth]{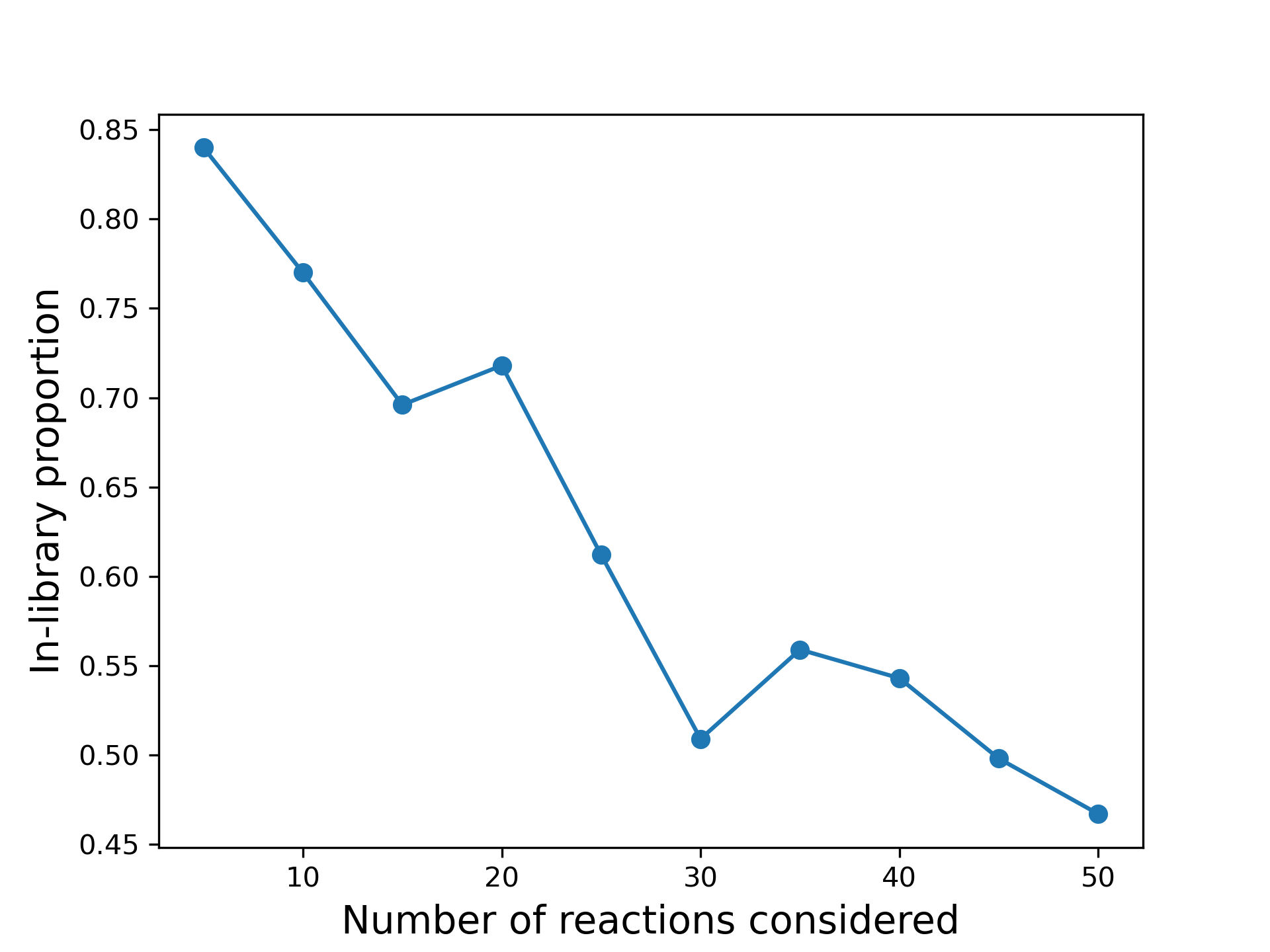}
    \caption{As the number of distinct reactions present in training increases, the in-library proportion for generated samples from RationaleRL quickly decreases, dropping to below 50\% after just 50 reactions are included.}
    \label{fig:extrapolation}
\end{figure}

\subsection{JT-VAE details}
JT-VAE is a graph-based variational autoencoder which generates molecules according to a tree-structured scaffold of fragments. Unlike RationaleRL, JT-VAE samples chemical \emph{fragments} in an autoregressive fashion, rather than atoms or bonds. Prior to training, we apply the vocabulary generation step described in the JT-VAE paper to the sampled products from the subset of REAL utilized in the baseline experiments, which yields a total of 325 fragments. Our experiments are based on the author-provided code, which can be found at \href{https://github.com/wengong-jin/icml18-jtnn}{\texttt{https://github.com/wengong-jin/icml18-jtnn}}. 

\subsection{CSLVAE details}

During training we utilize an annealing schedule on $\beta$ \cite{sonderby2016ladder} (see Algorithm \ref{alg:training}), starting with $\beta=0$ and incrementing by 1e-5 every 2000 iterations, with a max value of $\beta=1$. We train for a total of 200K iterations, in which time CSLVAE has seen a total of 2000 $\times$ 200K = 400M compounds (although not 400M unique compounds, due to the batch sampling strategy). As such, by the time training is halted, CSLVAE has seen no more than 2.5\% of the full REAL library.

\section{Test-time distribution shift in CSLVAE}

As described in Algorithm \ref{alg:training}, we sample small subsets of the full library in each training iteration for tractability. In particular, the library sub-sampler we utilize first samples uniformly over the reactions and subsequently samples a constant number of products in each reaction at random; the synthons associated with these sampled products comprise the synthons in the library subset. In training the CSLVAE model used in our experiments, we sample 20 reactions and 100 products per reaction, which yields a minibatch of 2000 products. As described in the paper, these associated library subsets describe a chemical space of roughly 300K-1.5M compounds each. At test time, however, we wish to decode according to the full library of 16B compounds, which constitutes a fairly drastic test-time distribution shift.

\begin{figure}
    \centering
    \includegraphics[width=0.8\linewidth]{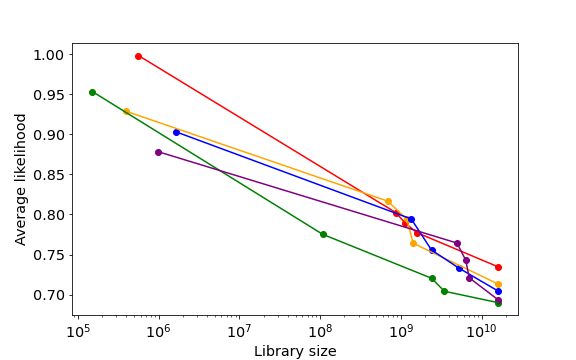}
    \caption{Test-time distribution shift as a function of library size.}
    \label{fig:testshift}
\end{figure}

Supplementary Figure \ref{fig:testshift} illustrates the magnitude of the test-time distribution shift on reconstruction quality (measured by average likelihood). We start by sampling five library subsets as described above, drawing 2000 compounds for each subset; each of these five draws is represented by a different color in the figure. We then grow these library subsets by first including all synthons across the 20 sampled reactions, and calculate the average reconstruction likelihood for the same 2000 products according to the expanded library, and further increase the library by adding all synthons for 100 randomly sampled additional reactions, and then all synthons for yet another 100 randomly sampled additional reactions, and finally include all synthons and all reactions. At each step, the library subsets describe an increasingly larger chemical space. We observe that the average likelihood decreases in a roughly linear-log fashion in the library size as a result of this distribution shift.

\section{Ex-post density estimation}

In Section 3.3 of the paper, we describe the use of ex-post density estimation on latent codes corresponding to products sampled uniformly from the library as a way to force random samples from CSLVAE to track more closely to sampling uniformly at random from the library. Algorithm \ref{alg:expost} outlines this procedure.

We carry out a simple experiment to demonstrate that this procedure achieves the intended result. The experiment proceeds as follows. We draw 10,000 molecules from the library uniformly at random which we treat as a training set for the ex-post density estimator. We consider three density estimators of increasing expressivity: a multivariate normal (MVNormal), a mixture of five normals (MoG-5), and a mixture of ten normals (MoG-10). We also compare to the standard approach of sampling from the isotropic multivariate normal prior. As such, we have four alternative schemes for sampling latent codes which are subsequently decoded into compounds from the library. We then sample another 10,000 molecules from the library, again uniformly at random, which is treated as a reference set for comparison.

\begin{table}[ht]
\centering
\caption{Divergences between query sampling strategies and a reference set of compounds sampled uniformly at random from the library.}
\begin{tabular}{l| c c c c c }
\hline
 & Train & Prior & MVNormal & MoG-5 & MoG-10 \\
\hline
Uniqueness & 100\% & 96.9\% & 99.8\% & 99.7\% & 99.8\% \\
SA (JSD) & 0.0179 & 0.3340 & 0.0651 & 0.0557 & 0.0452 \\
QED (JSD) & 0.0134 & 0.2200 & 0.0572 & 0.0266 & 0.0225 \\
MW (JSD) & 0.0169 & 0.4370 & 0.0970 & 0.0650 & 0.0686 \\
logP (JSD) & 0.0144 & 0.0259 & 0.0643 & 0.0686 & 0.0703 \\
\hline
\end{tabular}
\label{tab:divergence}
\end{table}

Following \cite{polykovskiy2020molecular}, we compare the aforementioned sets of generated molecules to the reference set of molecules on the following computable molecular properties: synthetic accessibility (SA), quantitative estimation of drug-likeness (QED), molecular weight (MW), and logarithm of the octanol-water partition coefficient (logP). In particular, we calculate the Jensen-Shannon distance (square root of the Jensen-Shannon divergence) between the distribution of these properties on the reference set and each set in question. Supplementary Table \ref{tab:divergence} summarizes the results of this exercise. Three items worth noting are that (a) the reference compounds and the compounds sampled for training the ex-post density estimators have low divergence across the various properties, (b) sampling from the prior generates compounds with high divergence relative to uniform sampling over the library, and (c) using more expressive density estimators for the latent codes leads to increasingly lower divergence across the various properties with the reference set, as they are able to better match the distribution of latent codes for the training set (which is exchangeable with the reference set).

\section{Comparison to existing analogue enumeration approaches}

We attempt a comparison of CSLVAE's analoguing capabilities with that of Arthor, a state-of-the-art commercial similarity search tool for synthesis libraries developed by NextMove Software. Arthor performs analogue enumeration using a custom ECFP4 bit vector representation of molecules, returning compounds with high Tanimoto similarity according to this fingerprint. For a given query compound, we enumerate the top-100 analogues returned by Arthor from REAL. Similarly, we encode each query compound with the CSLVAE encoder and generate a corresponding 100 stochastic decodings. For every analogue, we compute its RDKit ECFP4 Tanimoto similarity with the query compound and retain only the top-1 analogue. We then compare the distribution of Tanimoto similarities of the top-1 analogues returned in this way between Arthor and CSLVAE. As a control, we use a naive random baseline policy which samples 100 compounds at random from REAL, again selecting the top-1 analogue based on RDKit ECFP4 Tanimoto similarity. Because both CSLVAE and the random baseline constitute stochastic policies, we repeat this procedure 30 times for each query compound and take the average top-1 Tanimoto similarity.

\begin{table}[ht]
\centering
\caption{List of FDA novel drug approvals in 2021 used in analogue comparison.}
\begin{tabular}{l | l}
\hline
Molecule & Canonical SMILES \\
\hline
asciminib & {\tiny O=C(Nc1ccc(OC(F)(F)Cl)cc1)c1cnc(N2CCC(O)C2)c(-c2ccn[nH]2)c1} \\
atogepant & {\tiny CC1C(c2c(F)ccc(F)c2F)CC(NC(=O)c2cnc3c(c2)CC2(C3)C(=O)Nc3ncccc32)C(=O)N1CC(F)(F)F} \\
avacopan & {\tiny Cc1ccc(NC(=O)C2CCCN(C(=O)c3c(C)cccc3F)C2c2ccc(NC3CCCC3)cc2)cc1C(F)(F)F} \\
belumosudil & {\tiny CC(C)NC(=O)COc1cccc(-c2nc(Nc3ccc4[nH]ncc4c3)c3ccccc3n2)c1} \\
belzutifan & {\tiny CS(=O)(=O)c1ccc(Oc2cc(F)cc(C\#N)c2)c2c1C(O)C(F)C2F} \\
cabotegravir & {\tiny CC1COC2Cn3cc(C(=O)NCc4ccc(F)cc4F)c(=O)c(O)c3C(=O)N12} \\
casimersen & {\tiny CN(C)P(=O)(OC[C@@H]1CNC[C@H](n2ccc(=N)nc2O)O1)N1CCN(C(=O)OCCOCCOCCO)CC1} \\
drospirenone & {\tiny CC12CCC3c4ccc(O)cc4CCC3C1C(O)C(O)C2O} \\
finerenone & {\tiny CCOc1ncc(C)c2c1C(c1ccc(C\#N)cc1OC)C(C(N)=O)=C(C)N2} \\
fosdenopterin & {\tiny Nc1nc(=O)c2c([nH]1)N[C@@H]1O[C@@H]3COP(=O)(O)O[C@@H]3C(O)(O)[C@@H]1N2} \\
maralixibat & {\tiny CCCCC1(CCCC)CS(=O)(=O)c2ccc(N(C)C)cc2C(c2ccc(OCc3ccc(C[N+]45CCN(CC4)CC5)cc3)cc2)C1O} \\
maribavir & {\tiny CC(C)Nc1nc2cc(Cl)c(Cl)cc2n1[C@H]1O[C@@H](CO)[C@H](O)[C@@H]1O} \\
mobocertinib & {\tiny C=CC(=O)Nc1cc(Nc2ncc(C(=O)OC(C)C)c(-c3cn(C)c4ccccc34)n2)c(OC)cc1N(C)CCN(C)C} \\
piflufolastat & {\tiny O=C(O)CCC(NC(=O)NC(CCCCNC(=O)c1ccc(F)nc1)C(=O)O)C(=O)O} \\
ponesimod & {\tiny CCCN=C1SC(=Cc2ccc(OCC(O)CO)c(Cl)c2)C(=O)N1c1ccccc1C} \\
samidorphan & {\tiny NC(=O)c1ccc2c(c1O)C13CCN(CC4CC4)C(C2)C1(O)CCC(=O)C3} \\
serdexmethylphenidate & {\tiny COC(=O)C(c1ccccc1)C1CCCCN1C(=O)OC[n+]1cccc(C(=O)NC(CO)C(=O)[O-])c1} \\
sotorasib & {\tiny C=CC(=O)N1CCN(c2nc(=O)n(-c3c(C)ccnc3C(C)C)c3nc(-c4c(O)cccc4F)c(F)cc23)C(C)C1} \\
tepotinib & {\tiny CN1CCC(COc2cnc(-c3cccc(Cn4nc(-c5cccc(C\#N)c5)ccc4=O)c3)nc2)CC1} \\
tivozanib & {\tiny COc1cc2nccc(Oc3ccc(NC(=O)Nc4cc(C)on4)c(Cl)c3)c2cc1OC} \\
trilacicilib & {\tiny CN1CCN(c2ccc(Nc3ncc4cc5n(c4n3)C3(CCCCC3)CNC5=O)nc2)CC1} \\
umbralisib & {\tiny CC(C)Oc1ccc(-c2nn(C(C)c3oc4ccc(F)cc4c(=O)c3-c3cccc(F)c3)c3ncnc(N)c23)cc1F} \\
vericiguat & {\tiny COC(=O)Nc1c(N)nc(-c2nn(Cc3ccccc3F)c3ncc(F)cc23)nc1N} \\
viloxazine & {\tiny CCOc1ccccc1OCC1CNCCO1} \\
\hline
\end{tabular}
\label{tab:fda2021}
\end{table}

\begin{figure}
    \centering
    \includegraphics[width=0.8\linewidth]{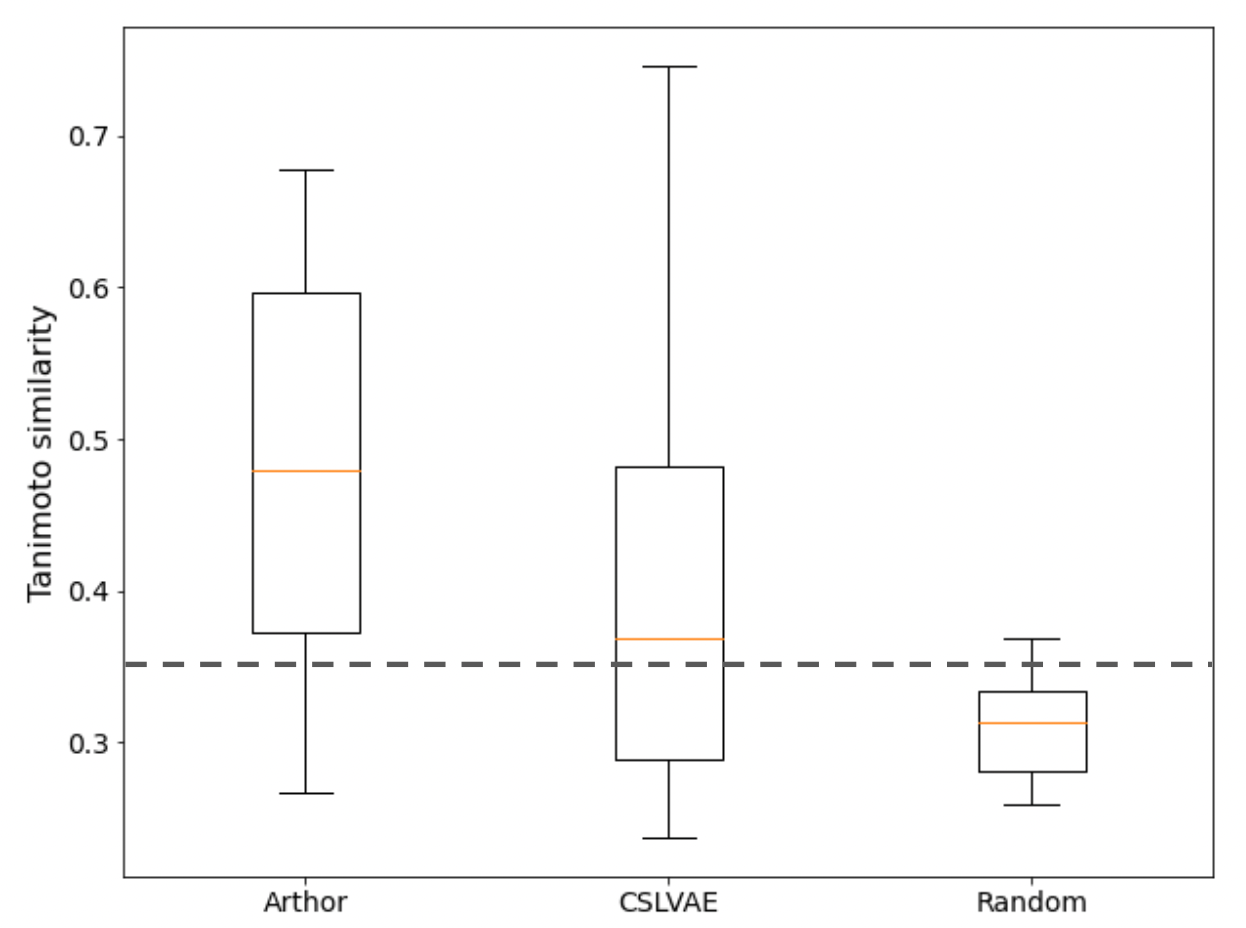}
    \caption{Comparing CSLVAE to Arthor in top-1 analogue search.}
    \label{fig:arthor}
\end{figure}

For the query compounds, we use 24 of the 51 novel drugs approved in 2021 by the FDA, filtering out drugs which do not satisfy a routine set of small molecule criteria (such as monoclonal antibodies); the compounds used in this experiment are listed in Supplementary Table \ref{tab:fda2021}. Supplementary Figure \ref{fig:arthor} shows a boxplot of the top-1 Tanimoto similarities for Arthor, CSLVAE, and the random baseline. While CSLVAE finds more distant ECFP4 analogues compared to Arthor (which is a gold standard for fingerprint similarity search), it is nonetheless able to identify analogues\footnote{It is typical to use a Tanimoto similarity threshold in the range 0.3-0.35 to indicate whether a pair of molecules can be seen as analogues.} for unseen, novel drugs in a routine manner.

Of particular note, CSLVAE is considerably less resource intensive than Arthor, which is specially designed for fast and efficient fingerprint-based analogue enumeration and requires appropriate infrastructure and setup. For these reasons, it is difficult to do a direct apples-to-apples comparison of the compute requirements. Nevertheless, we share the following pertinent information. The CSLVAE experiments were run on a machine with an NVIDIA Tesla K80 GPU and an Intel Xeon E5-2686 CPU. On average, sampling 100 analogues for a given query using CSLVAE completed in 11.41 seconds with this setup (encoding and decoding). Further, all of the parameters and buffers (including representations and keys for the synthon, R-group, and reactions) of the trained CSLVAE model required just 170MB memory. By comparison, the Arthor experiments are run in a distributed fashion on 100 pods, each with 8 CPUs, with top-100 analogue enumeration requiring a total of 32.46 seconds, for an approximate total CPU time of 25,968 seconds. Furthermore, Arthor needs to devote rather significant amounts of memory and storage to carry out analogue enumeration; our in-house setup uses 3GB per shard and makes 64GB RAM requests for each Arthor worker.

We note that CSLVAE's ability to represent large CSLs with significantly fewer resources and perform analogue retrieval with notably improved execution time (perhaps three orders of magnitude faster) owes to its decoding strategy, which utilizes parallel synthon look-ups, thereby requiring a number of keys that is on the order of number of synthons in the library rather than on the order of number of products in the library, and further permits a kind of similarity search in time that is logarithmic in the number of products in the library (rather than linear).

As a final remark, we view this exercise as a demonstration of CSLVAE's potential as a backbone in search applications, not as a definitive implementation of an analogue enumeration strategy with CSLVAE. There are a number of requirements of a good analogue enumeration tool that we purposefully do not focus on here (e.g., enumerating $k$ distinct analogues rather than top-1 analogue, where $k$ could be large). While this may be of interest in subsequent work, it is not our focus in this paper. We believe that the quality of analogues found with CSLVAE-based decoders can begin to approach those found by existing, specially-tailored tools for analogue enumeration with additional efforts aimed at such applications, and that such directions could prove fruitful in an era of combinatorial library explosion that challenges traditional enumerative strategies.

\section{Encoder transfer to molecular property prediction}

The CSLVAE training objective can be viewed as a kind of contrastive pretext task, which seeks to align the representation of a given molecule with representations that correspond to retrieval instructions in a CSL. Hence, it is natural to wonder whether the encoder of a trained CSLVAE model could demonstrate good transfer performance in prediction tasks that may be of interest. To investigate, we compare MLPs trained on CSLVAE with those trained on molecular fingerprints (ECFP4 and ECFP6) for molecular property prediction tasks. We use the octanol-water partition coefficient (logP) and the quantitative estimate of drug likeness (QED) \cite{bickerton2012quantifying} as targets for prediction.

\begin{table}[ht]
\centering
\caption{Encoder transfer on logP and QED prediction. The cells report the average RMSE ± one standard deviation, calculated over five runs.}
\begin{tabular}{c | c c c c c c}
\hline
	& & REAL 100K	& ZINC 250K & REAL 100K & ZINC 250K \\
	& Dimensionality & logP & logP & QED & QED \\
\hline
CSLVAE	& 64	& \textbf{0.539 ± 0.002}	& 0.591 ± 0.001	& \textbf{0.072 ± 0.001}	& 0.068 ± 0.001 \\
ECFP4	& 256	& 0.827 ± 0.001	& 0.679 ± 0.002	& 0.091 ± 0.002	& 0.079 ± 0.001 \\
ECFP6	& 1024	& 0.601 ± 0.002	& \textbf{0.490 ± 0.001} & \textbf{0.072 ± 0.001} & \textbf{0.064 ± 0.001} \\
\hline
\end{tabular}
\label{tab:contrastive}
\end{table}

For this experiment, we construct a dataset by sampling 100K compounds uniformly at random from REAL, splitting the examples into training, validation, and testing folds using an 80-10-10 split. For each compound, we extract as feature descriptor its CSLVAE query, in addition to its ECFP4 and ECFP6 fingerprints. Using the training fold, we fit an MLP on each of these feature descriptors separately to predict the molecule's logP and QED score. We select the iteration which attains the lowest validation RMSE, recording its test RMSE. This is repeated five times, and we report the average test RMSE and standard deviation. To demonstrate the extent to which CSLVAE learns molecular features that are predictive of such molecular properties for out-of-domain compounds, we repeat this exercise on a dataset of 250K molecules from ZINC. The results of this exercise are summarized in Supplementary Table \ref{tab:contrastive}.

The results of this experiment confirm that the latent space learned by CSLVAE can indeed be utilized to success in predicting quantities like logP and QED, especially in the in-domain case where it out-performs the predictors fit on chemical fingerprints. However, in the out-of-domain case, the predictor fit on ECFP6 fingerprints performs notably better than the predictor fit on CSLVAE queries, suggesting that the features learned by CSLVAE may be missing some pertinent predictive information about input molecules which differ significantly from the CSL on which it was trained.

\end{document}